\documentclass[%
twocolumn,
 amsmath,amssymb,
 aps, physrev,
]{revtex4-2}

\usepackage{graphicx}
\usepackage{dcolumn}
\usepackage{bm}
\usepackage{hyperref}

\usepackage{float}
\usepackage{units}
\usepackage{color}

\begin{document}

\title{\textbf{Enhanced long-duration gravitational-wave transient sources search pipeline with denoising and tree clustering algorithms} 
}

\author{Hugo Einsle}
\email{Contact author: hugo.einsle@oca.eu}
\affiliation{Universit\'e C\^ote d'Azur, Observatoire de la C\^ote d'Azur, CNRS, Laboratoire Artemis, 06300 Nice, France}
 
\author{Marie Anne Bizouard}
\affiliation{Universit\'e C\^ote d'Azur, Observatoire de la C\^ote d'Azur, CNRS, Laboratoire Artemis, 06300 Nice, France}

\author{Adrian Macquet}
\affiliation{
Universit\'e Paris-Saclay, CNRS/IN2P3, IJCLab, 91405 Orsay, France}

\affiliation{ }

\date{\today}

\begin{abstract}
We present a two–stage upgrade to the \textsc{PySTAMPAS} pipeline that boosts the search for long–duration (\unit[$10$--$10^{3}$]{s}) transients in gravitational-wave detectors' data. First, a denoising scheme combines complex 2-D wavelet shrinkage with adaptive pixel threshold to suppress noise while retaining signal power. Second, a KDTree nearest-neighbour algorithm clusters surviving pixels in $\mathcal{O}(\log(n))$ time, replacing the standard clustering approach. Tests with a week of LIGO O3b data show a large reduction in false alarm rate and an improvement of the search sensitivity by up to a factor 2. Moreover, the computational time has been significantly reduced. These gains extend the sensitivity of all-sky, all-time searches to weaker and shorter transients, paving the way for rapid and deeper analyses in forthcoming LIGO-Virgo-KAGRA observation campaigns.
\end{abstract}


\maketitle

\section{\label{sec:introduction} Introduction}
The first detection of gravitational-wave emission from the merger of two black holes~\citep{LIGOScientific:2016aoc} started the field of gravitational-wave astronomy. Many compact binary mergers have since been discovered~\citep{LIGOScientific:2018mvr, LIGOScientific:2020ibl, LIGOScientific:2021usb,KAGRA:2021vkt} in ground-based gravitational-wave detectors. Yet, other transient gravitational-wave sources are expected to be discovered thanks to the sensitivity increase of the LIGO~\citep{Aasi:2014pky}, Virgo~\citep{Acernese:2014yos} and KAGRA~\citep{Aso:2013eba} detectors forming the LVK network of detectors. Along with the detectors' sensitivity increase, transient gravitational-wave search algorithms can be improved to detect weak signals buried in the noise. Among these gravitational-wave detection pipelines \texttt{PySTAMPAS}~\citep{Macquet:2021ttq, pySTAMPAS} has been developed to detect transient gravitational-wave signals whose duration can last up to several hundreds of seconds in the frequency band of LVK detectors, coming from a specific sky position (target mode) or from any position in the sky (all-sky mode). Signals are expected from compact object mergers~\citep{Abbott:2018urg}, isolated neutron stars~\citep{Andersson:1997rn} or the collapse of the core of massive stars that form a black hole with an accretion disk when the inner layers of the star fall back onto the compact remnant. Long duration (\unit[$\mathcal O (10-100)$]{s}) gravitational waves may be emitted by disk turbulence and disk instabilities fragmentation~\citep{Piro:2006ja}.Long duration transients have been searched in LVK network data ~\cite{Abbott:2017vtl, Abbott:2019laf, Abbott:2021bhs, Abbott:2018urg, Abbott:2022sts} with the Coherent WaveBurst~\cite{Klimenko:2015ypf}, STAMP-AS~\cite{Abbott:2015kgh} and X-SphRad~\cite{Cannon:2007br} search pipelines.

\texttt{PySTAMPAS} is a hierarchical pipeline designed for analyzing year-long dataset from multiple detectors, inspired from the STAMP library~\cite{Thrane:2010ri} and its initial STAMP-AS implementation~\cite{Abbott:2015kgh}. It first constructs signal-to-noise ratio spectrograms ($ft$-maps) for each individual detector, where the presence of a gravitational-wave signal appears as an excess of power within pixels. To select signals, a threshold is applied to the $ft$-map pixels. A seed-based clustering algorithm such as \texttt{burstegard}~\citep{Prestegard:2016, Prestegard:2012} is used to group pixels based on their proximity in time and frequency, to form signal clusters (stage 1). The thresholding step implies that low amplitude signal pixels below threshold are lost for the cross-correlation second stage where gravitational-wave signal are reconstructed (stage2). Moreover, \texttt{burstegard} is computationally bounded when the number of pixels becomes large which happens when the data are particularly noisy. Alternative clustering algorithms such as \texttt{Seedless}~\citep{Thrane:2013bea} have been proposed. They do not require to select pixels above a threshold but match simple track patterns in $ft-$maps. These algorithms are efficient for smooth signal morphologies but can be computationally limited in case of large dataset to analyze~\citep{Thrane:2014bma}.

The efficacy of \texttt{PySTAMPAS} has been demonstrated through its application to the data since the LVK O3 observing run~\citep{Macquet:2021eyn, Macquet:2021xue}. Yet, the clustering stage as implemented in \texttt{PySTAMPAS} has been identified as a bottleneck to quickly isolate pixels containing gravitational-wave signals. 
Improved clustering could enable us to analyze smaller clusters, enhancing sensitivity to shorter-duration signals in the \unit[$1-10$]{s} range, rather than the current focus on longer durations of \unit[$10-1000$]{s}. Additionally, this improvement could open up the possibility of extending the search to higher-frequency bands ($>$\unit[2]{kHz}), which are often overlooked due to noise contamination and computational loads. In this context, denoising plays a critical role as a preparatory step for clustering by reducing the noise contamination.

To improve the detection performance of unmodelled gravitational-wave transients of a wide range of duration and frequency, we propose a novel clustering algorithm for \texttt{PySTAMPAS} that integrates denoising techniques via wavelet analysis, adaptive thresholding, and \texttt{KDTree}-based clustering~\citep{Bentley:1975}. This clustering algorithm was developed for efficient multidimensional search and query operations primarily in computer science and its applications, particularly in spatial data analysis and nearest neighbor searches~\citep{Friedman:1977}. 

In Section~\ref{sec:methods} we first summarize how data from multiple detectors are processed by \texttt{PySTAMPAS}. We then describe the methodology developed for data denoising via wavelet analysis, adaptive thresholding and clustering with \texttt{KDTree}. Section~\ref{sec:performance} reports the performance of the new \texttt{PySTAMPAS} features considering simulated Gaussian dataset as well as the LVK data from O3. We then draw some conclusions in Section~\ref{sec:conclusion}. 

\section{\label{sec:methods} PySTAMPAS detection algorithm improvements}

\subsection{\label{sec:denoising} $ft$-maps denoising with adaptive threshold}

\texttt{PySTAMPAS} single-detector $ft$-maps are given by 
\begin{equation}
\tilde{y}(t;f) = \frac{\tilde{s}(t;f)}{\sqrt{P(t;f)}}\, ,
\end{equation}
where $\tilde{s}(t;f)$ is the spectrogram of the gravitational-wave strain time-serie $s(t)$\, obtained using the one-sided Fourier transform over short segments of \unit[1]{s} and $P(t;f)$ is the noise power spectral density (PSD) of each segment (see~\citep{Macquet:2021eyn} for details).
Long-duration gravitational-wave signals are expected to spread over tens to hundreds of seconds. We thus focus on removing noise pixels arising from noise fluctuations that persist over scales smaller than a few seconds or a few hertz. For this purpose, we combine a wavelet denoising step with an adaptive threshold.

The 2D wavelet transform (WT) provides an effective framework for analyzing signals with both frequency and temporal variations~\citep{meyer1993wavelets,mallat2008wavelet}, scaling a 2D mother wavelet function $ \psi_a(u, v) $. More precisely, the 2D WT of the function $ \tilde{y}(t, f) $ is given by

\begin{equation}
W(a, \tau, \nu) = \iint \tilde{y}(t, f)\, \psi_{a}\left(\frac{t - \tau}{a}, \frac{f - \nu}{a}\right)\, dt \, df \, ,
\end{equation}

where the parameter \( a \) represents the scale factor, adjusting the wavelet's width. When \( a > 1 \), the wavelet is stretched to capture larger-scale signal components, while \( a < 1 \) compresses the wavelet, capturing smaller-scale signal components. The scaled wavelet function is expressed as \( \psi_{a}(u, v) = \frac{1}{|a|} \psi \left(\frac{u}{a}, \frac{v}{a}\right) \).
The central position parameters \( \tau \) and \( \nu \) are translation factors in the time and frequency domains, respectively. They shift the wavelet across the signal, allowing localized analysis of different regions of \( \tilde{y}(t, f) \). 


To implement the 2D WT, we use the PyWavelets library~\citep{Lee2019} which can handle complex inputs. When using a real wavelet filter (as is the case in this study), the 2D WT is applied separately to the real and imaginary parts of $\tilde{y}(t,f)$.  The output of the transformation thus consists of complex-valued approximation and detail coefficients, representing different scale components of the $ft$-map. This wavelet transformation is applied across both dimensions of the $ft$-map, effectively dividing the coefficients into four complex-valued sub-bands (LL, LH, HL, HH). Each sub-band captures distinct structural information within the $ft$-map, aiding in the separation of signal and noise.

The LL sub-band, also known as the approximation coefficients, captures the large-scale content of the $ft$-map. This component acts as a smoothed version of the $ft$-map, emphasizing continuous changes over extended time and frequency ranges. In the context of gravitational-wave detection, LL coefficients retain essential information about long-duration signals, representing the overall energy distribution across the map.
The LH sub-band contains horizontal detail coefficients, capturing small-scale variations in frequency that persist over larger time scales. These coefficients are sensitive to fine frequency variations within the $ft$-map and may capture noise or transient signal components with rapid frequency changes that persist across time segments.
The HL sub-band contains vertical detail coefficients, emphasizing small-scale variations in time while remaining continuous across frequency. This sub-band captures fine temporal variations and can indicate transient events or bursts of power that occur briefly in time but span broader frequency ranges. Noise components that are localized in time yet spread across multiple frequency bins may appear in this sub-band.
Finally, the HH sub-band captures diagonal detail coefficients, which emphasize highly localized variations in both time and frequency. This sub-band is particularly responsive to small-scale, high-energy fluctuations that do not extend broadly over time or frequency. In $ft$-maps, HH coefficients often correspond to noise fluctuations, making them useful for selective denoising by removing them. We denote the set of wavelet coefficients as \( W_{i}^{\alpha} \), where \( \alpha \) represents one of the sub-bands (LL, LH, HL, HH). 

To eliminate the pixels of noise, we apply a threshold on the detail coefficients \( W_{i}^{\text{HH}} \). 
We draw randomly $N$ $W_{i}^{\text{HH}}$ coefficients and compute the standard deviation of the sample
\begin{equation}
  \sigma_W = \sqrt{\frac{1}{N} \sum_{i=1}^{N} \left(W_{i}^{\text{HH}} - \overline{W}^{\text{HH}}\right)^2}\, ,
  \label{eq:sigma_W} 
\end{equation}
where $ \overline{W}^{\text{HH}} $\, is the mean of the $N$ $W_{i}^{\text{HH}}$\, coefficients. This calculation is repeated across $10$ randomly chosen samples, and the median of $\sigma_W$\, values is used to define the threshold $\Sigma_W$\, on the detail coefficients. More precisely, we apply soft-thresholding to the $W_{i}^{\text{HH}}$\, coefficients

\begin{equation}
W_{i}^{*\text{HH}} =
\begin{cases}
W_{i}^{\text{HH}} - \Sigma_W & \text{if } W_{i}^{\text{HH}} > \Sigma_W \\
0 & \text{otherwise}\, ,
\end{cases}    
\end{equation}
where \( W_{i}^{*\text{HH}} \) denotes the thresholded HH coefficients. This soft-thresholding approach removes small pixel structures resulting from noise fluctuations while preserving significant features from gravitational-wave signals and continuity. After thresholding, a two-dimensional inverse wavelet transform is performed to reconstruct the denoised $ft$-map, denoted as $\tilde{y}^*(t, f)$.

We then apply an adaptive threshold to the pixels of the denoised $ft$-maps $\tilde{y}^*(t,f)$ to further isolate significant features. 
Since the denoised map $\tilde{y}^*(t,f)$ is complex-valued, we base the subsequent thresholding steps on the magnitude (absolute value) of the pixel values. 
For each denoised $ft$-map, we form $10$ subsets $S^j$\, of $m$ pixels randomly drawn without replacement. 
For each subset, we compute the standard deviation of the pixel magnitudes
\begin{equation} 
\sigma_{|\tilde{y}^*|}^j = \sqrt{\frac{1}{m} \sum_{i=1}^{m} \left( |S_i^j| - \overline{|S|}^j \right)^2}\, , 
\label{eq:sigma_mag_y_star} 
\end{equation}
where $|S_i^j|$\, represents the magnitude of the complex value of the $i$-th pixel in sample $S^j$, and $\overline{|S|}^j$\, is the mean of these magnitudes. 
For each sample, the threshold value $\phi^j$\, is given by $\sigma_{|\tilde{y}^*|}^j$\, scaled with the square root of twice the natural logarithm of the sample size $m$, as suggested by Donoho and Johnstone~\citep{donoho1994ideal}

\begin{equation}
   \phi^j = \sigma_{|\tilde{y}^*|}^j \sqrt{2 \ln(m)}\, . 
\end{equation}

The adaptive threshold $\Phi$, defined as the median of the $\phi^j$ values, is then applied to the denoised $ft$-map $\tilde{y}^*(t, f)$

\begin{equation}
|\tilde{y}^*(t, f)| > \alpha  \Phi\, ,    
\end{equation}
where $\alpha$ is a scaling factor to tune the threshold. Calculating an adaptive threshold for each $ft$-map ensures that the thresholding process is tailored to the specific characteristics of the data. Utilizing the median minimizes the influence of large amplitude pixels associated with potential gravitational-wave signals, ensuring that the threshold accurately reflects the underlying noise level.

\subsection{\label{sec:clustering} KDTree clustering}

\texttt{KDTree} is a multi-dimensional binary search trees algorithm which has been implemented as a data structure for storage of information to be retrieved by associative searches~\citep{Bentley:1975}. When used for nearest neighbor queries its running-time performance, scaling as ${\cal O}(log(n))$, outperforms standard algorithms. We use the algorithm for the case of two dimensions or axes corresponding to the time and frequency axes of the multi-resolution $ft$-map. Pixels that pass the adaptive threshold (referred to \textit{seeds}) are grouped into clusters using a \texttt{KDTree} data structure~\citep{Maneewongvatana:1999, 2020SciPy-NMeth}.

The \texttt{KDTree} is constructed recursively by partitioning the set $P$ of seeds of coordinates ($p_t$, $p_f$) along alternating axes. Initially, seeds are partitioned along the time axis according to their position with respect to the median $\theta_t$\, of their $p_t$ coordinates such that we obtain two new subsets
\begin{align}
    P_{\text{left}} &= \{p \in P \mid p_t \leq \theta_t\}\, , \nonumber \\
    P_{\text{right}} &= \{p \in P \mid p_t > \theta_t\}\, .
\end{align}

The next partition is performed along the frequency axis such that, if we call the iteration number the depth ($d$), we know which axis is used to build the partition at depth $d$
\begin{equation}
\text{axis}(d)=
\begin{cases}
\text{time} & \text{if } d \text{ is even}\\
\text{frequency} & \text{if } d \text{ is odd}\, .
\end{cases}
\end{equation}

At depth $d$, each parent subset $P(d)$ of seeds is split into two child subsets, creating a \textit{branch}
\begin{align}
P_{\text{left}}(d+1) &= \{p \in P(d) \mid p_{\text{axis}(d)} \leq \theta(d)\}\, , \nonumber \\
P_{\text{right}}(d+1) &= \{p \in P(d) \mid p_{\text{axis}(d)} > \theta(d)\}\, .    
\end{align}

\begin{figure}[h]
\includegraphics[width=.5\textwidth]{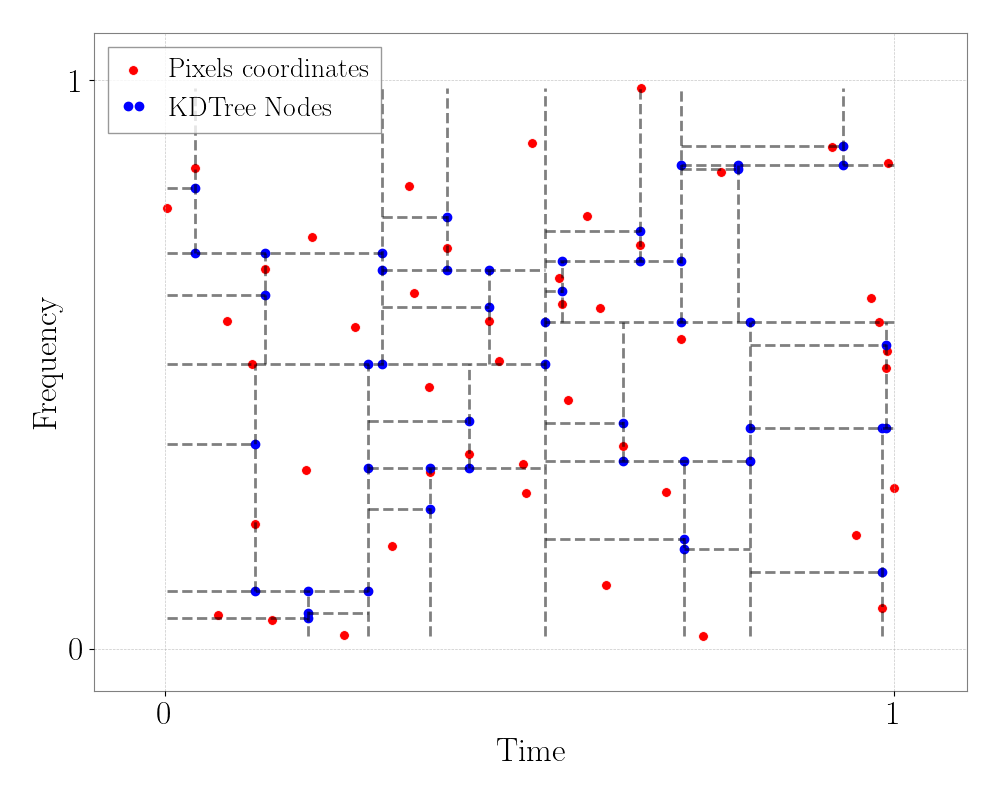}
\caption{Sixth iteration (depth 5) in the construction of the \texttt{KDTree} containing 46 seeds. The horizontal and vertical dashed lines are branches. Nodes are in blue. A branch connected to only one node is a leaf.}
\label{fig:KDTree_depth5}
\end{figure}

Figure \ref{fig:KDTree_depth5} shows an example of a KDTree that is completed after six iterations.

Once the \texttt{KDTree} is built, a range search algorithm is used to identify neighboring seeds that should be grouped into clusters. For a given query seed $p$, the algorithm finds all seeds $p'$ that fall within a specific normalized distance $r$, the search radius.
For a couple of seeds $p$ and $p'$, the normalized distance metric is defined as
\begin{equation}
\delta(p, p') = \sqrt{\left(\frac{|p_t - p'_t|}{\Delta t}\right)^2 + \left(\frac{|p_f - p'_f|}{\Delta f}\right)^2}\, ,
\label{eq:kdtree_distance} 
\end{equation}
where $\Delta t$ and $\Delta f$ are the time and frequency resolutions of the $ft$-map, respectively. The neighboring seed \(p'\) is considered part of the same cluster as \(p\) if
\begin{equation}
\delta(p, p') \leq r\, .
\label{eq:kdtree_condition} 
\end{equation}




The \texttt{KDTree} structure enables efficient pruning during the range search. At each internal node, the algorithm checks if the search region defined by the query seed $p$ and radius $r$ intersects the splitting hyperplane defined by the node's splitting axis and value. If the search region lies entirely on one side of the hyperplane, the algorithm does not need to explore the subtree corresponding to the other side, significantly reducing the number of distance calculations compared to a brute-force approach.

Once all neighboring seeds of a query seed $p$ within radius $r$ are identified, they are grouped into a cluster. The algorithm iteratively expands each cluster by performing range searches for neighbors of newly added seeds, continuing until no further seeds satisfy the proximity condition. This process results in multiple clusters, potentially including clusters containing only a single seed. Analysis requirements typically impose a minimum cluster size.



    


\section{Performance and comparison}
\label{sec:performance}

\subsection{Dataset and configuration}
\label{sec:dataset}
We evaluate the performance of the new denoising and clustering features of \texttt{PySTAMPAS} by comparing its detection capabilities with respect to the standard \texttt{burstegard} algorithm~\citep{Macquet:2021eyn}. This comparison is done using LIGO data from the O3b observing run. The chosen data consists of seven days of coincident data from the LIGO detectors Hanford (LHO) and Livingston (LLO), segmented into 1,181 windows, each \unit[512]{s} long, with an overlap of 50\%~\citep{KAGRA:2023pio}. The search covers a large frequency band from \unit[22]{Hz} to \unit[2]{kHz}.
For each detector, $ft$-maps are constructed. The PSD of the  $ft$-maps is estimated by taking the median of the squared modulus of the Fourier transform over the \unit[512]{s} segment considered~\citep{Macquet:2021ttq}.
The list of the search parameters is provided in Table~\ref{tab:parameters}.
The clustering radius of the \texttt{burstegard} algorithm is set to \unit[2]{s} in time and \unit[2]{Hz} in frequency. This corresponds to a setting used in the most recent search for long-duration transient gravitational-wave sources in LIGO-Virgo-KAGRA data~\citep{Abbott:2021bhs,LIGOScientific:2025ksg}. We also empirically set the \texttt{KDtree} radius to $2$. We require a minimum cluster size of 50 pixels for both \texttt{burstegard} and \texttt{KDTree}. These two parameters are directly related to the morphological properties of the targetted signals.
Finally, in Section~\ref{sec:optimization} we explain how the \texttt{DAT} parameters have been chosen.

\begin{table}[h]
\centering
\begin{tabular}{ll}
\hline
\textit{$ft$-maps} & \\
Detectors & LHO and LHO \\
Window duration & \unit[512]{s} \\
Frequency range & \unit[$22 - 2000$]{Hz} \\
Time-frequency resolution & \unit[4.0]{s} $\times$ \unit[0.25]{Hz}, \unit[2.0]{s} $\times$ \unit[0.5]{Hz},\\
 & \unit[1.0]{s} $\times$ \unit[1.0]{Hz}, \unit[0.5]{s} $\times$ \unit[2.0]{Hz} \\
 PSD estimation & full-median \\

\hline
\textit{Denoising} \texttt{DAT} &\\
$N$ & $7.4 \times 10^{5}$\\
$m$ & $10^5$\\
$\alpha$ & 0.85\\

\hline
\textit{Clustering} \texttt{KDTree} &\\
$r$ & 2 \\
Minimal number of pixels & 50\\

\hline
\textit{Clustering} \texttt{burstegard} &\\
Pixel energy & 2\\
Radius & \unit[2]{s} in time and \unit[2]{Hz} in frequency \\
Minimal number of pixels & 50\\

\hline
\textit{Coherent stage} & \\
Ranking statistic & $p_\Lambda$\\
\hline

\end{tabular}
\caption{Values of the \texttt{PySTAMPAS} parameters used for the performance comparison of \texttt{DAT+KDTree} and \texttt{burstegard} algorithms.}
\label{tab:parameters}
\end{table}

\begin{figure}[h]
    \centering
    \includegraphics[width=.45\textwidth]{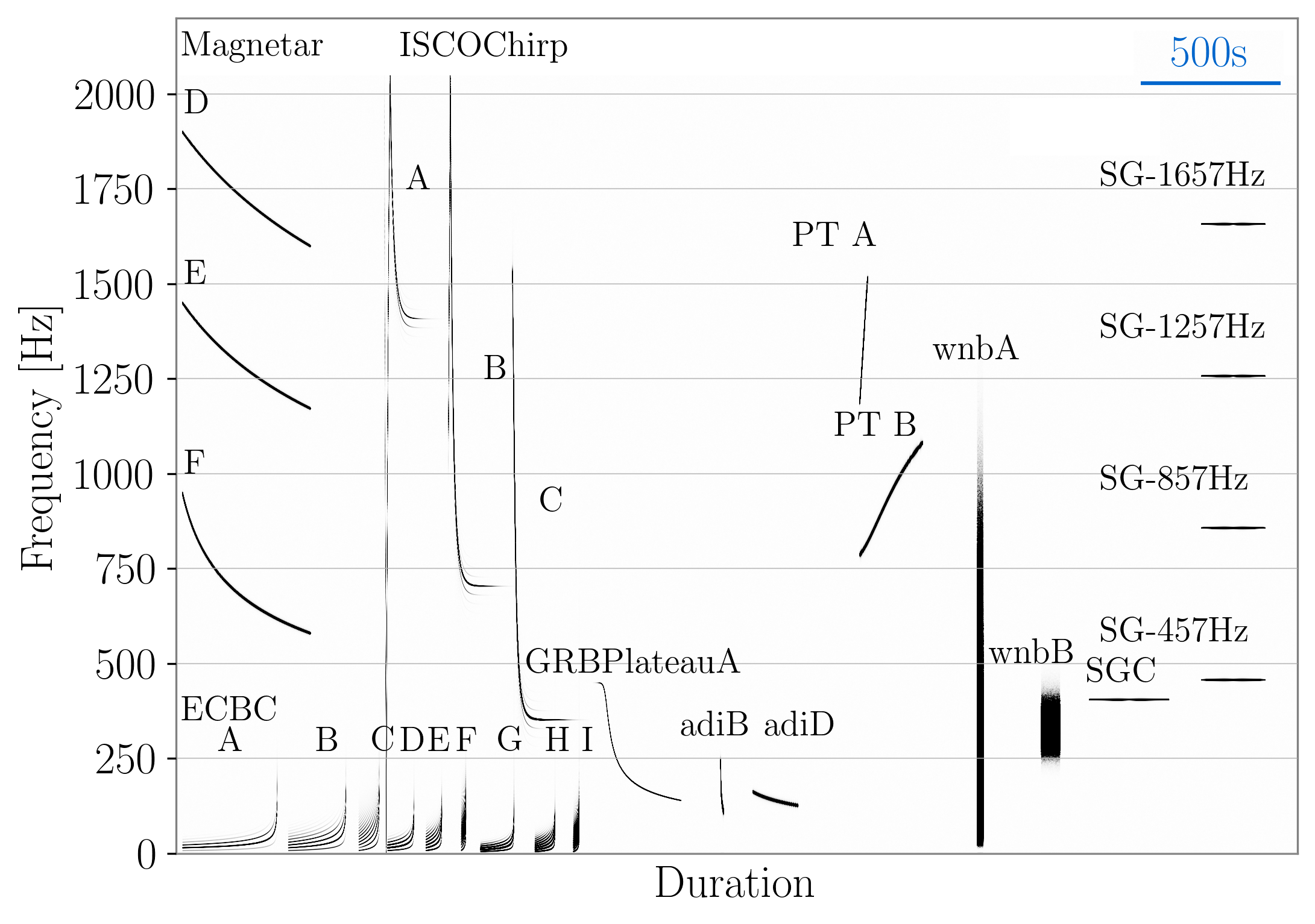}
    \caption{Time-frequency spectrogram of all waveforms used to estimate the detection efficiency of PySTAMPAS in this article.}
    \label{fig:waveforms}
\end{figure}

To estimate the detection efficiency of \texttt{PySTAMPAS}, we consider several transient gravitational-wave signal waveforms to span the large parameter space in time and frequency. The astrophysical gravitational-wave models include post-merger magnetar (\texttt{magnetar})~\citep{2015ApJ...798...25D}, black hole accretion disk instabilities (\texttt{ADI})~\citep{PhysRevLett.87.091101}, newly formed magnetar powering a gamma-ray burst plateau (\texttt{GRBplateau})~\citep{Corsi_2009}, eccentric inspiral-merger-ringdown compact binary coalescence waveforms (\texttt{ECBC})~\citep{2018PhRvD..97b4031H}, fallback accretion on a neutron star (\texttt{PT})~\citep{2012ApJ...761...63P}, and broadband chirps from innermost stable circular orbit waves around rotating black holes (\texttt{ISCOchirp})~\citep{van_Putten_2016}. They are completed with ad-hoc waveforms such as white noise burst (\texttt{WNB}), sine-Gaussian (\texttt{SG}) and damped-Sine (\texttt{DS}). We vary the signal amplitude such that the detection efficiency is well sampled. All other parameters, including arrival time, sky position, cosine of the inclination angle, and polarization angle, are uniformly randomized.
Figure~\ref{fig:waveforms} shows a time-frequency representation of the signal's waveforms, highlighting the distinct spectral evolution characteristic of each astrophysical source and the wide parameter space covered.

\subsection{\label{sec:optimization} \texttt{DAT} parameters optimization}

The \texttt{DAT} algorithm has few parameters that need to be optimized to optimally detect gravitational-wave signals in LVK data. In addition to the choice of the wavelet transform, the algorithm has three main parameters: $N$ the number of HH detail coefficients used to compute the threshold applied in the 2D WT domain, $m$ the number of denoised pixels to compute the threshold on the pixel amplitude and $\alpha$ the scaling factor applied to this threshold.
Because optimization is highly dependent on the targeted gravitational-wave signal and the properties of the data noise, we have not tried to optimize all parameters altogether. In the following, we provide indications about the range of values of the parameters that optimize the algorithm performance for a wide variety of gravitational-wave signal morphologies. 

\begin{figure}[h]
    \centering
    \includegraphics[width=0.5\textwidth]{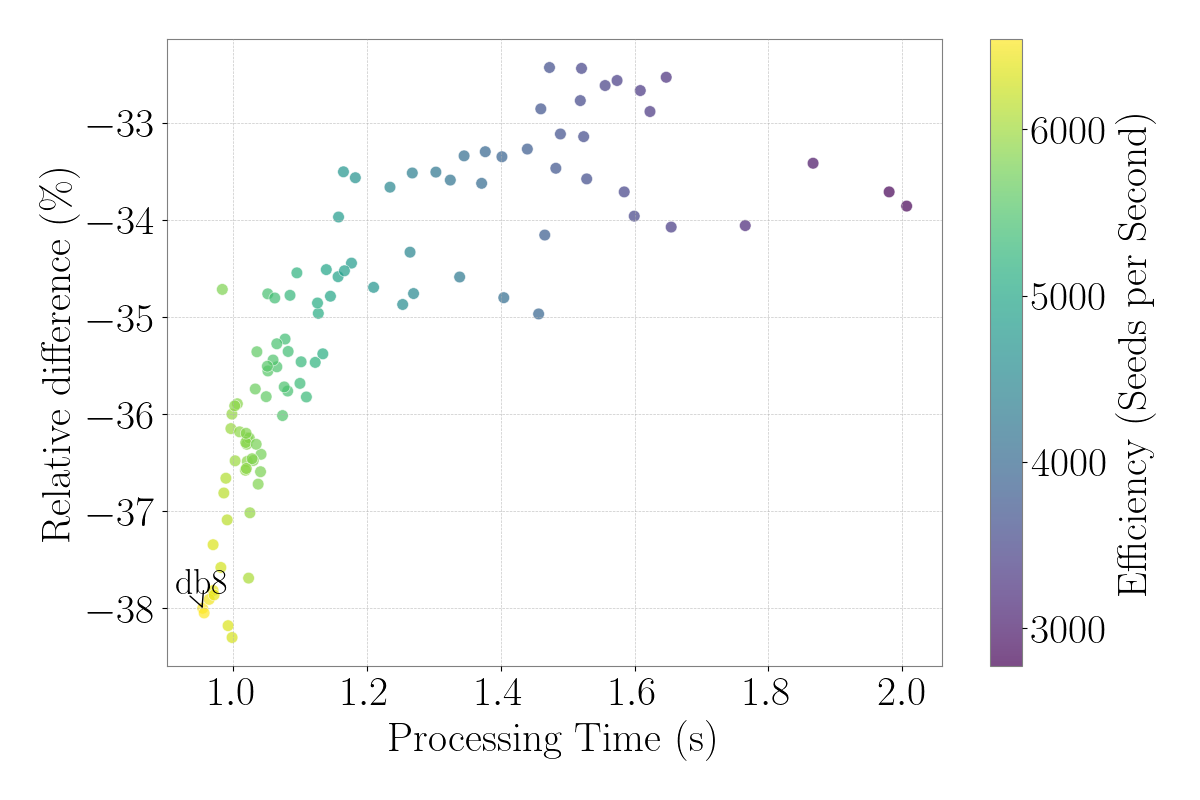}
    \caption{Relative difference of the number of $ft$-map pixels due to denoising as function of the processing time of the wavelet transform. Each marker correspond to one of the 106 wavelet transforms available in the \texttt{PyWavelets} library. The color of the marker indicates the number of seeds decrease per second of processing time.}
    \label{fig:DATwavelet_histo}
\end{figure}

First, we choose the wavelet transform of the \texttt{DAT} algorithm from all the 106 transforms available in the \texttt{PyWavelets} library~\cite{Lee2019}. We  consider the denoising performance of a $ft$-map built from gravitational-wave synthetic data reproducing the LIGO Hanford detector noise during the O3b LVK Observing run~\citep{O3bsensitivity:2019}. More precisely, we compare the noise pixels reduction due to denoising for each of the wavelet transforms. The other two parameters $N$ and $m$ are chosen such that we select 10\% of the $W_i^{\rm HH}$ coefficients and 10\% of the denoised $ft$-map pixels respectively. This corresponds to $N \approx 7.5 \times 10^4$ coefficients of each $ft$-map resolution, and $m \approx 10^5$ pixels for each $ft$-map resolution.
We build \unit[512]{s} $\times$ \unit[20-2000]{Hz} $ft$-maps including four resolutions (\unit[4]{s} $\times$ \unit[0.25]{Hz}, \unit[2]{s} $\times$ \unit[0.5]{Hz}, \unit[1]{s} $\times$ \unit[1]{Hz}\, and \unit[0.5]{s} $\times$ \unit[2]{Hz}).  To compare $ft$-maps, we consider pixels with absolute amplitude larger than $2$. This threshold applied to original and denoised $ft$-maps correspond to a selection of $2-5\%$ pixels in the non-denoised maps. This allows to focus on the highest amplitude noise pixels which impact the most searches. We then average, over $20$ $ft$-maps, the relative difference of the number of pixels for each of the 106 transforms. 
One of the wavelet transform, the Reverse Biorthogonal 3.1, is excluded as it removed 100\% of the seeds, rendering it ineffective for signal detection. To select the optimal wavelet, we consider the ratio between the pixels number reduction and the processing time taken by the wavelet transform (efficiency). Figure~\ref{fig:DATwavelet_histo} shows the results for the 105 wavelets, where the markers' color indicates the efficiency of each wavelet transform.
The Daubechies 8 (db8) wavelet, highlighted in the bottom-left region of the figure, stands out as the most efficient wavelet. It achieves one of the highest seed removal fraction (approximately 38\%) within the minimum processing time.

We now aim at optimizing the $\alpha$ parameter that controls the denoising adaptive threshold. To do so, we add gravitational-wave signals to the simulated noise and for a given value of $\alpha$\, we compute the number of signal pixels above the threshold (True Positives or $TP$), the number of signal pixels below the threshold (False Negatives or $FN$), the number of noise pixels above the threshold (False Positives or $FP$) and the number of noise pixels below the threshold (True Negatives or $TN$). This allows to compute the sensitivity $S$ and the specificity $P$ defined as
\begin{equation}
S = \frac{TP}{TP+FN} \quad \text{and} \quad P = \frac{TN}{TN+FP}\, ,
\label{eq:sensitivity_specificity_def} 
\end{equation}
These quantities are averaged over 20 independent noise realizations to account for noise fluctuation and the sample sizes $N$ and $m$ are fixed to the same values than before. 

To make the optimization more robust, we use several gravitational-wave signals among those described in Section~\ref{sec:dataset}: ADI\_A, magnetar\_D, magnetar\_E, ISCOchirp\_A and ECBC\_C waveforms supplemented by a broad range of ad-hoc DS signals with central frequency between $55$ and \unit[$1620$]{Hz} and duration between $1$ and \unit[$50$]{s}\footnote{The letter suffix in the waveform's name refer to some parameters whose definition is given in~\cite{Macquet:2021xue}}. Each waveform is added to the noise data with different amplitudes for which we compute sensitivity $S$ and specificity $P$.
We also compute the objective function defined as $\mathcal{F} = \frac{S \times P}{S + P}$. 

\begin{figure}[h]
    \centering
    \includegraphics[width=.45\textwidth]{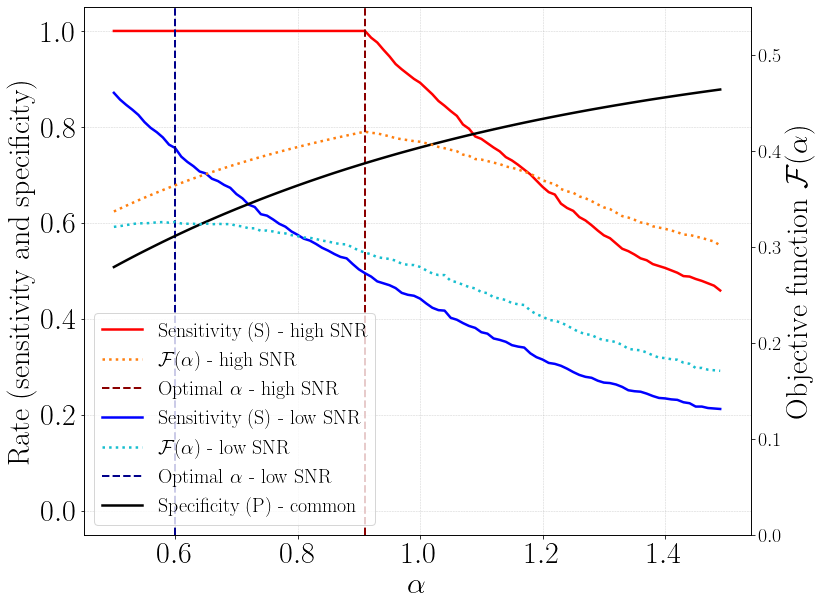}
\caption{
Sensitivity and specificity of the denoising adaptive thresholding algorithm as function of the scaling factor $\alpha$ of the pixel amplitude threshold for two cases of signal's amplitude (red for high SNR signals and blue for low SNR signals). The dashed line curves show the objective functions.
The red dashed vertical line marks the maximum objective value for a scaling factor of $\alpha = .91$ for high SNR signals. The blue dashed vertical line corresponds to the maximum objective value ($\alpha = .61$) for the low SNR case.}
    \label{fig:sensitivity_specificity}
\end{figure}

Figure~\ref{fig:sensitivity_specificity} shows the value of the sensitivity, specificity and objective as a function of the parameter $\alpha$ for added signals averaged over all types of gravitational-wave signal for two values of signals' amplitude, corresponding to large detection efficiency (high SNR) and the other one to signals poorly detected (low SNR).
Low values of $\alpha$ leads to high sensitivity by capturing all signal pixels, but at the cost of reduced specificity, as more noise pixels are selected. As $\alpha$ increases, the specificity improves, correctly rejecting more noise pixels. Sensitivity shows a constant decline as more signal pixels fail to pass the threshold.
The optimum for the high SNR signals given by the maximum of the objective function is achieved for $\alpha=.91$. This corresponds to the detection of $\sim$ 100\% of signal pixels while $\sim$ 72\% of noise pixels are rejected. When we decrease the amplitude of the signals from maximal detectability (high SNR) to minimal detectability (low SNR), the optimal value of $\alpha$ decreases from $.91$ down to $.61$. Choosing a small value of $\alpha$ is adapted when one looks for low amplitude signal buried in the noise. The drawback is the high specificity which could be an issue in terms of computational time and data management when large volume of data need to be processed. In other words, such a low $\alpha$ might be adapted in case of a \textit{targeted search} for which the volume of data to analyze is limited around the time of the event. On the other hands, for a search of several months of data, it is recommended to choose larger $\alpha$ value.

\begin{figure}[t]
    \centering
    \includegraphics[width=.45\textwidth]{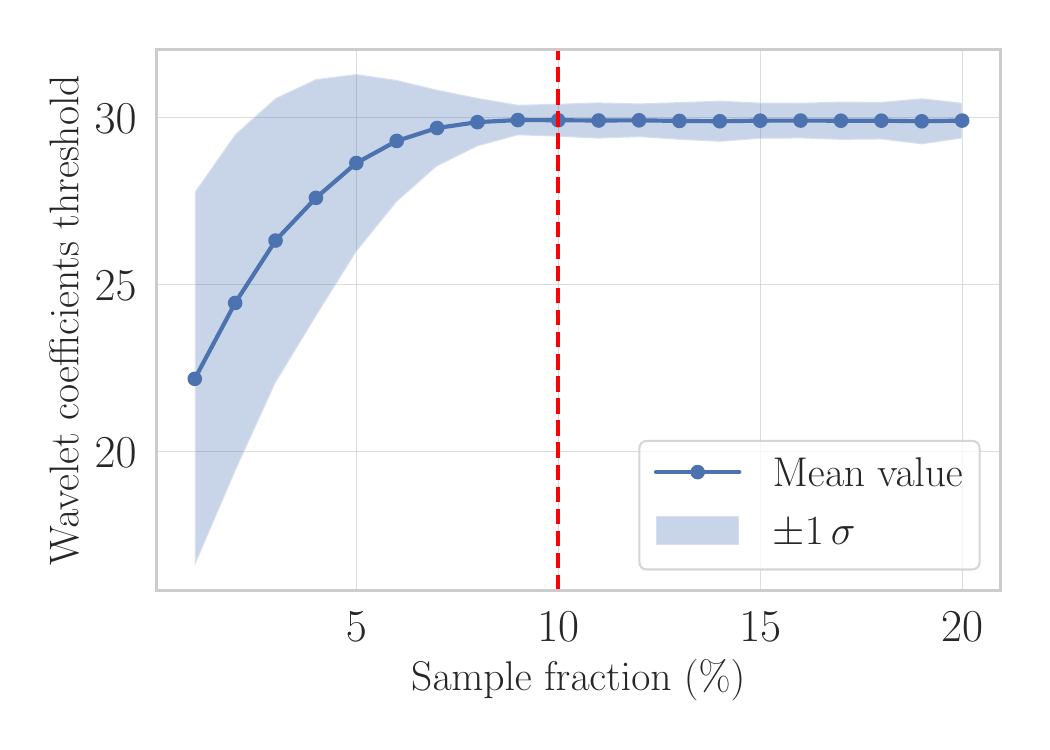}
     \includegraphics[width=.45\textwidth]{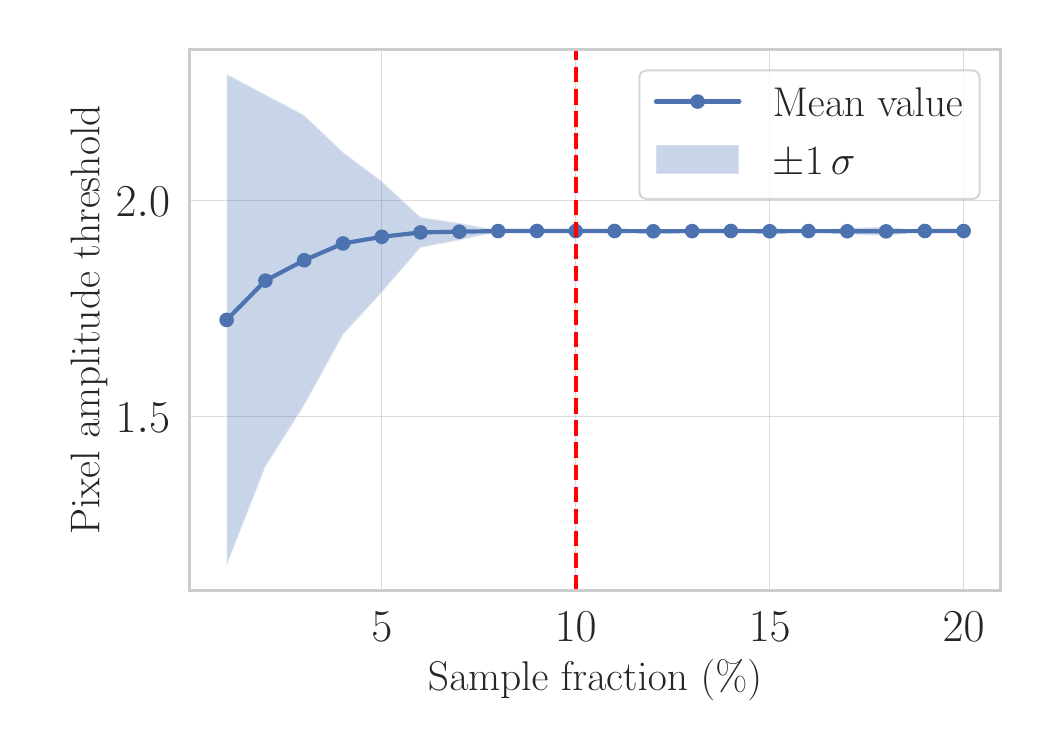}
    \caption{
    Top panel: wavelet coefficients threshold ($\Sigma_W$) as a function of the fraction of HH wavelet coefficients. 
    Bottom panel: pixel amplitude threshold ($\alpha \Phi$) as a function of the pixels number fraction.
    The blue solid line is the mean value across the 30 $ft$-maps and the shaded region correspond to $\pm1\sigma$. The red dashed line points out the fraction of sample for which the threshold value has converged.
    }
    \label{fig:denoising_convergence}
\end{figure}

Finally, we study the dependence of the wavelet coefficient threshold $\Sigma_W$ and the adaptive pixel amplitude threshold $\alpha \Phi$ with respect to the number of detail coefficients $N$ and the number of pixels $m$ used to estimate the two thresholds.
To do that, we consider data from the LVK O3b run to take into account non Gaussian features of the detector noise. We generate 30 $ft$-maps from randomly chosen times within the six months of O3b run and compute $\Sigma_W$\, and $\alpha \Phi$.
In Figure~\ref{fig:denoising_convergence}, we show $\Sigma_W$ and $\alpha \Phi$ as functions of the fraction of the number of HH detail coefficients and the fraction of the number of pixels, respectively. In both cases, 
the thresholds are confidently estimated for a fraction greater than 10\%.

The optimisation of the \texttt{DAT} parameters depends on the statistical properties of the data and the type of search carried out. Yet, we have found that $N$, and $m$, can be fixed to 10\%, regardless of the amount of data and signals. On the other hand the choice of the $\alpha$ parameter depends on the amount of data to analyse. Searching for a gravitational-wave signal in a limited amount of data allows us to consider a low $\alpha$ threshold to mine weak signals into the noise because the excess of noisy pixels should be easily handled in the next stages of the analysis. In the case of a search covering a full year of data, an excessive number of noise clusters could cause difficulties, both in terms of data management and computational time.
In the following section, we consider $\alpha=.85$ which is a threshold adapted to a search of a gravitational-wave signal in a large data set (\textit{all-sky/all-time} search).

\subsection{Results}
\subsubsection{Noise-only analysis}
 

The incoherent stage (stage 1) of \texttt{PySTAMPAS} is applied to the 1-week O3b data from the LHO and LLO detectors. When $ft$-maps are built we mask the high-amplitude spectral features corresponding to known instrumental artefacts present in real interferometer data~\cite{Abbott:2021bhs}. The standard \texttt{burstegard} algorithm extracts 10\% more clusters from LHO and 36\% more from LLO compared to \texttt{DAT+KDTree}. In both detectors, \texttt{burstegard} shows overall a significantly higher cluster density, but both algorithms show excesses especially at low frequency (below \unit[100]{Hz} where the anthropogenic and interferometer control noises dominate).
For each cluster found in stage 1, we compute the coherent trigger using the other detector $ft-$map pixels (stage 2). To artificially increase the effective duration of observation we apply a time-shift between the two data streams. Sliding the data many times allows us to increase the effective observing time while keeping the real data properties. The time shift is large enough ($>$\unit[2]{s}) to suppress any coherence between the two data streams. In this study, we perform 64 slides of \unit[8]{s} in each analysis window and we shift 75 times the other data stream analysis windows, resulting in a total of 4800 time-shifts which is equivalent to 50.5 years of signal-free data. The coherent triggers are ranked using the $p_\Lambda$ detection statistic~\citep{Macquet:2021ttq}. 

\begin{figure}[h]
    \centering
    \includegraphics[width=0.45\textwidth]{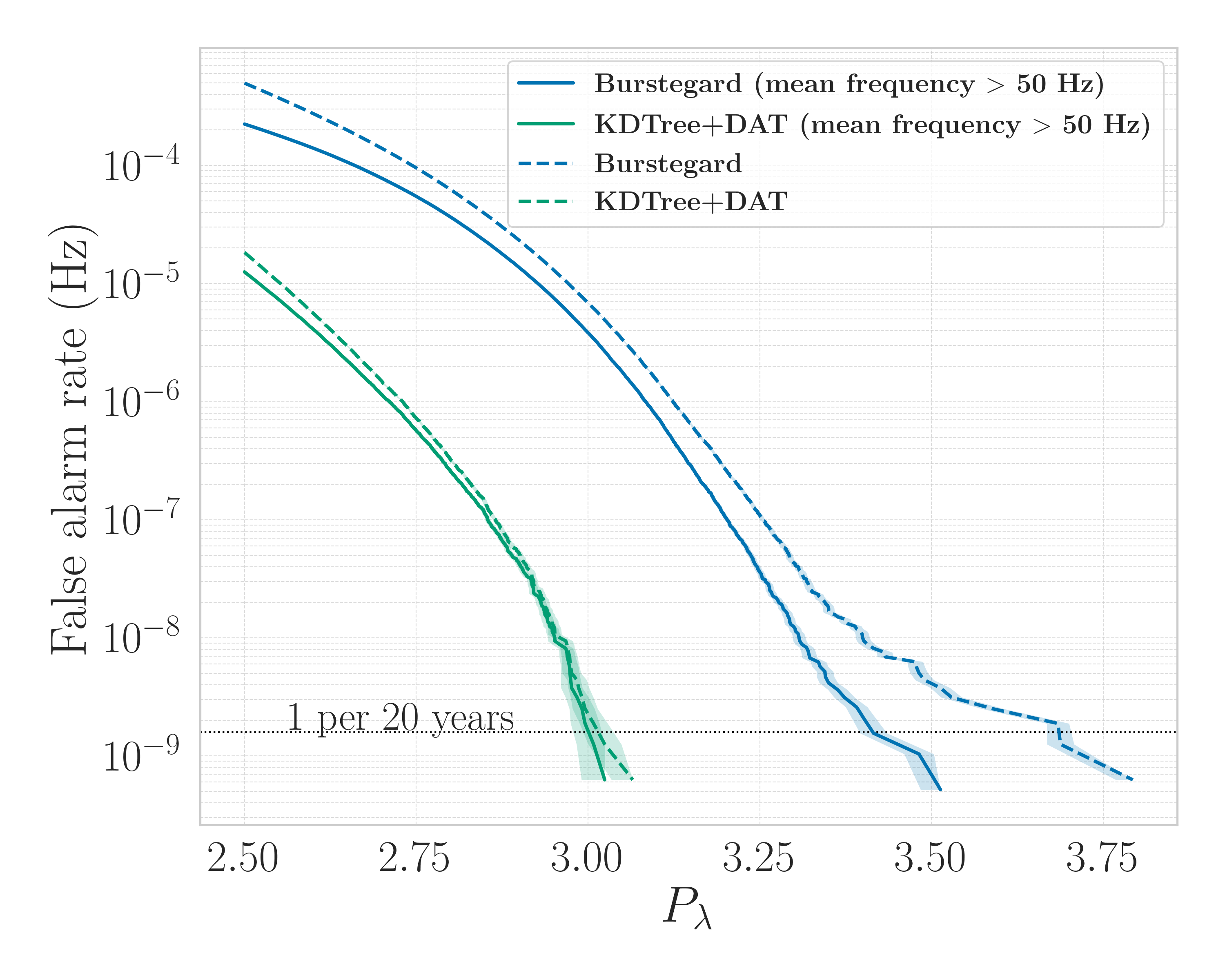}
    \caption{Normalized cumulative distributions of the noise triggers generated by \texttt{burstegard} (blue) and \texttt{DAT+KDTree} (green) algorithms using one week of LIGO Hanford and LIGO Livingston O3b data~\citep{KAGRA:2023pio} as function of their $p_\Lambda$ value. The distribution is normalized to the observation time. The shaded region is the 1-sigma uncertainty of a Poisson distribution.}

    \label{fig:burstegard_vs_KDTree_noise_estimation}
\end{figure}

Because of an excess of triggers at low frequency, affecting particularly \texttt{burstegard}, we have required that the mean frequency of the triggers be larger than \unit[50]{Hz}.
Figure~\ref{fig:burstegard_vs_KDTree_noise_estimation} compares the false alarm rate as a function of the coherent detection statistic $p_\Lambda$ for both algorithms. The overall false alarm rate is $\sim$ an order of magnitude lower for \texttt{DAT+KDTree}.
The false alarm rate reduction suggests that \texttt{DAT+KDTree} is efficient at disregarding noise pixels and is less sensitive to non-Gaussian transient noise events that usually populate the tail of the distributions. We have checked that \texttt{DAT} is almost entirely responsible for the false alarm reduction.

\subsubsection{Efficiency improvements}

To evaluate the detection efficiency of both the \texttt{burstegard} algorithm and \texttt{DAT+KDTree}, we inject randomly simulated signals into the same period of O3b data using the astrophysically-driven and ad-hoc models described in Section~\ref{sec:dataset}. 

The detection efficiency, defined as the fraction of injected signals that have $p_\Lambda$ larger than a given threshold, is estimated as function of the root-sum-square amplitude $h_\mathrm{rss}$ defined as 
\begin{equation}
    h_\mathrm{rss}= \sqrt{\int (h_+^2(t) + h_\times^2(t))dt}\, ,
\end{equation}
where $h_+(t)$ and $h_\times(t)$ are the two signal's polarizations in the source frame.
We have applied the same selection criteria than for the noise-only analysis (mean frequency $>$ \unit[50]{Hz}).
The detection efficiencies of both algorithms are evaluated at a false alarm rate of \unit[$1.5 \times 10^{-9}$] (one trigger in $20$ years). This corresponds to $p_\Lambda = 3.0$ for \texttt{DAT+KDTree} and $p_\Lambda = 3.41$ for \texttt{burstegard}.

\begin{figure}[htbp]
    \centering
    \includegraphics[width=0.45\textwidth]{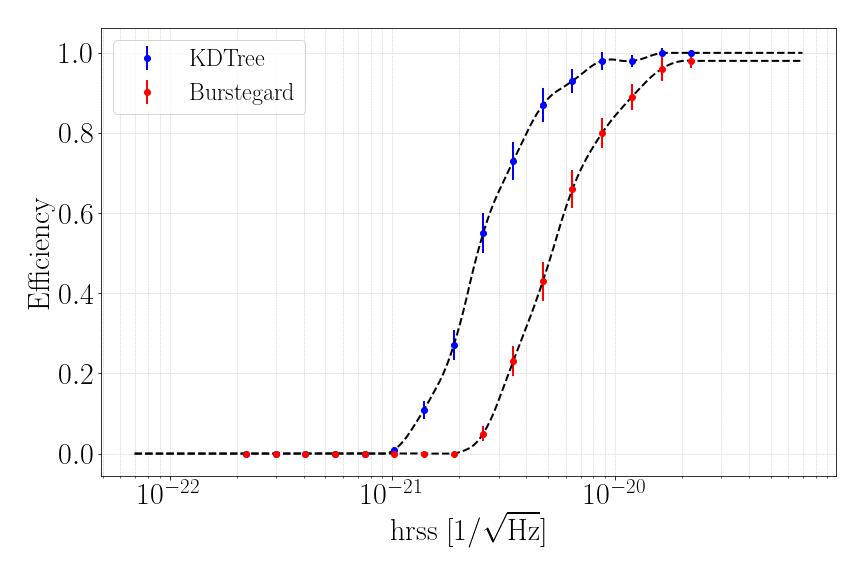}
    \caption{Example of a detection efficiency curve for the \texttt{magnetar\_E} signal, comparing \texttt{DAT+KDTree} (blue diamonds) and \texttt{burstegard} (red circles) for a false alarm rate of \unit[$1.58\times10^{-9}$]{Hz}. Efficiency is plotted as a function of the root-sum-square amplitude $h_{\mathrm{rss}}$. The efficiency error bars are computed assuming a binomial probability distribution of the number of recovered signals. A quadratic interpolation function is used to fit with the data points (dashed curves).}
    \label{fig:magnetarF_efficiency_example}
\end{figure}

Figure~\ref{fig:magnetarF_efficiency_example} shows an example of the detection efficiency as a function of $h_{\rm rss}$ for the \texttt{magnetar\_F} waveform, comparing the performance of \texttt{DAT+KDTree} with the \texttt{burstegard} algorithm. The continuous efficiency curves are obtained via a quadratic spline interpolation, which produces a smooth, piecewise polynomial curve that passes through each of the data points~\cite{LIGOScientific:2025ksg}.

In Table~\ref{tab:efficiencies} we report the root-sum-square amplitude $h_{\rm rss}$ corresponding to a detection efficiency of 50\%.
\texttt{DAT+KDTree} demonstrates higher detection efficiency compared to the \texttt{burstegard} algorithm for all waveforms. The gain is between a factor $1.0$ and $2.3$. 
The lowest gain, 5\%, is obtained for a fallback accretion on a neutron star waveform (\texttt{PT\_A}). For this waveform, and few others, their frequency range overlaps a large number of frequencies that are masked. We think the effect of these masks on a signal plays a dominant role compared to the denoising. On the other hands, high gain is obtained for different types of waveform.
This shows that the improvement barely depends on the precise morphology of the gravitational-wave signal, nor its frequency, nor its duration.

\begin{table*}[t]
\centering
\begin{tabular}{lcccc}
\hline
& & \multicolumn{1}{c}{\texttt{KDTree+DAT}} & \multicolumn{1}{c}{\texttt{Burstegard}} & \\
Waveform             &   & hrss@50\%              & hrss@50\%              & Ratio \\
\hline
adi\_B               &   & 4.0e-22               & 8.1e-22               & 2.0 \\ 
adi\_D               &   & 3.7e-22               & 4.4e-22               & 1.2 \\ 
ISCOchirpA           &   & 4.0e-21               & 5.0e-21               & 1.3 \\ 
ISCOchirp\_B         &   & 1.3e-21               & 2.7e-21               & 2.1 \\ 
ISCOchirp\_C         &   & 7.5e-22               & 1.4e-21               & 1.9 \\ 
GRBPlateau\_A        &   & 7.2e-22               & 1.1e-21               & 1.5 \\ 
PT\_A                &   & 1.3e-21               & 1.3e-21               & 1.0 \\ 
PT\_B                &   & 9.7e-22               & 1.2e-21               & 1.2 \\ 
ECBC\_A              &   & 1.1e-21               & 1.8e-21               & 1.6 \\ 
ECBC\_B              &   & 1.0e-21               & 1.6e-21               & 1.6 \\ 
ECBC\_C              &   & 6.8e-22               & 7.8e-22               & 1.1 \\ 
ECBC\_D              &   & 1.2e-21               & 1.9e-21               & 1.6 \\ 
ECBC\_E              &   & 1.1e-21               & 1.6e-21               & 1.5 \\ 
ECBC\_F              &   & 8.3e-22               & 1.3e-21               & 1.6 \\ 
ECBC\_G              &   & 1.7e-21               & 2.6e-21               & 1.5 \\ 
ECBC\_H              &   & 1.5e-21               & 2.2e-21               & 1.5 \\ 
ECBC\_I              &   & 1.1e-21               & 1.6e-21               & 1.5 \\ 
magnetarD            &   & 2.6e-21               & 3.5e-21               & 1.3 \\ 
magnetar\_E          &   & 2.4e-21               & 5.2e-21               & 2.2 \\ 
magnetar\_F          &   & 1.2e-21               & 2.5e-21               & 2.1 \\ 
WNB\_A               &   & 1.2e-21               & 2.0e-21               & 1.7 \\ 
WNB\_B               &   & 9.6e-22               & 1.7e-21               & 1.8 \\ 
SG\_C                &   & 6.2e-22               & 1.4e-21               & 2.3 \\ 
sg-200s-1657Hz       &   & 1.8e-21               & 2.2e-21               & 1.2 \\ 
sg-200s-1257Hz       &   & 1.1e-21               & 1.3e-21               & 1.2 \\ 
sg-200s-857Hz        &   & 6.7e-22               & 7.7e-22               & 1.1 \\ 
sg-200s-457Hz        &   & 4.0e-22               & 4.6e-22               & 1.1 \\ 
\hline
\end{tabular}
\caption{Comparison of $h_{\rm rss}$ values obtained at 50\% detection efficiency for \texttt{burstegard} and \texttt{DAT+KDTree} algorithms for different waveforms. Detection efficiency is estimated for a false alarm rate of \(1/20yr\). The ratio is defined as the $h_{\rm rss}$ of \texttt{burstegard} divided by the $h_{\rm rss}$ of \texttt{DAT+KDTree}.} 
\label{tab:efficiencies}
\end{table*}

\subsubsection{Signal reconstruction}

To further investigate the properties of the \texttt{DAT+KDTree} algorithm, we now compare the capability of each algorithm, \texttt{burstegard} and \texttt{DAT+KDTree}, to reconstruct a long-duration gravitational-wave signal. 
We add an \texttt{ADI\_D} signals into coincident LHO and LLO data segments of \unit[512]{s} from the O3b observing run. The amplitude and the frequency of the \texttt{ADI\_D} signal decrease with time.
The coherent analysis (stage 2) is performed using the clusters generated by each algorithm configuration with the settings given in Table~\ref{tab:parameters}.

\begin{figure}[h]
    \centering
    \includegraphics[width=.45\textwidth]{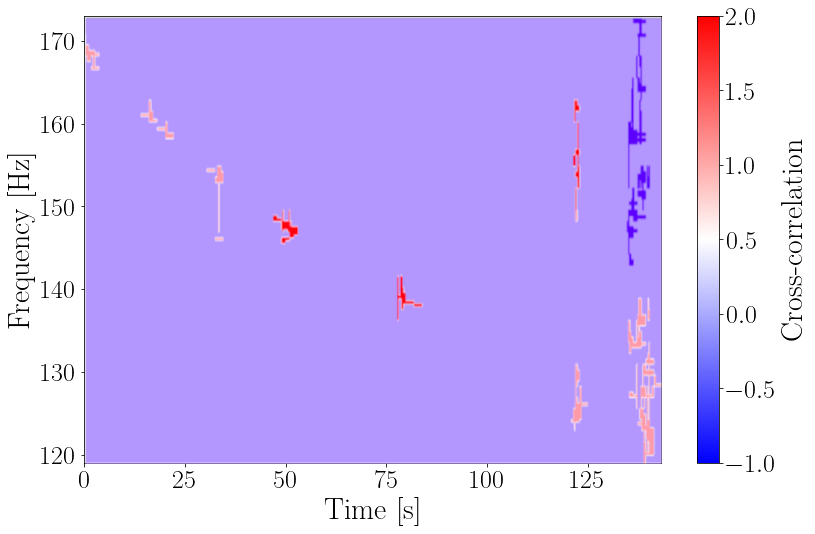}
    \caption{Superposition of the coherent pixels of an \texttt{ADI} waveform added in O3b LIGO data. The colors indicate which algorithm is recovering which part of the gravitational-wave signal: the bright red regions flag pixels identified by both algorithms, light red those exclusively found by \texttt{DAT+KDTree}, and in blue those exclusively found by \texttt{burstegard}. Notably, \texttt{DAT+KDTree} demonstrates a better recovery at the beginning and the end of the signal and disregards clusters associated with a noise artefact (glitch) that coincidences with the end of the signal.}
    \label{fig:signal_reconstruction_comparison}
\end{figure}

In Figure~\ref{fig:signal_reconstruction_comparison}, we overlay the coherent pixels identified by each algorithm for a given gravitational-wave signal: in light red, pixels selected only by the \texttt{DAT+KDTree}, in blue pixels found only by the \texttt{burstegard}, and in bright red, pixels found by both. 
Overall, \texttt{DAT+KDTree} identifies more pixels associated with the signal compared to \texttt{burstegard}. It selects signal pixels that are missed by \texttt{burstegard}, especially at the beginning and the end of the signal. 
Furthermore, the large glitch that almost coincides with the end of the gravitational-wave signal is less prominent in the \texttt{DAT+KDTree} result compared to \texttt{burstegard}. This improved discrimination against the glitch likely results from the combined effect of the \texttt{DAT} stage (reducing noise content before clustering) and the subsequent \texttt{KDTree} clustering method.
Specifically, \texttt{KDTree}'s ability, as described previously, to recursively partition the data along alternating dimensions using median splits ensures that clusters are more tightly bound in both time and frequency. This can improve the separation between genuine gravitational-wave signals and nearby localized noise artifacts, especially compared to \texttt{burstegard}.
In contrast, \texttt{burstegard}'s clustering approach may more readily group together noise artifacts that span larger regions of the time-frequency plane, leading to contamination of gravitational-wave signal clusters by nearby noise and higher false alarm rate.

\subsubsection{Execution time}

\begin{figure}[h]
    \centering
    \includegraphics[width=.45\textwidth]{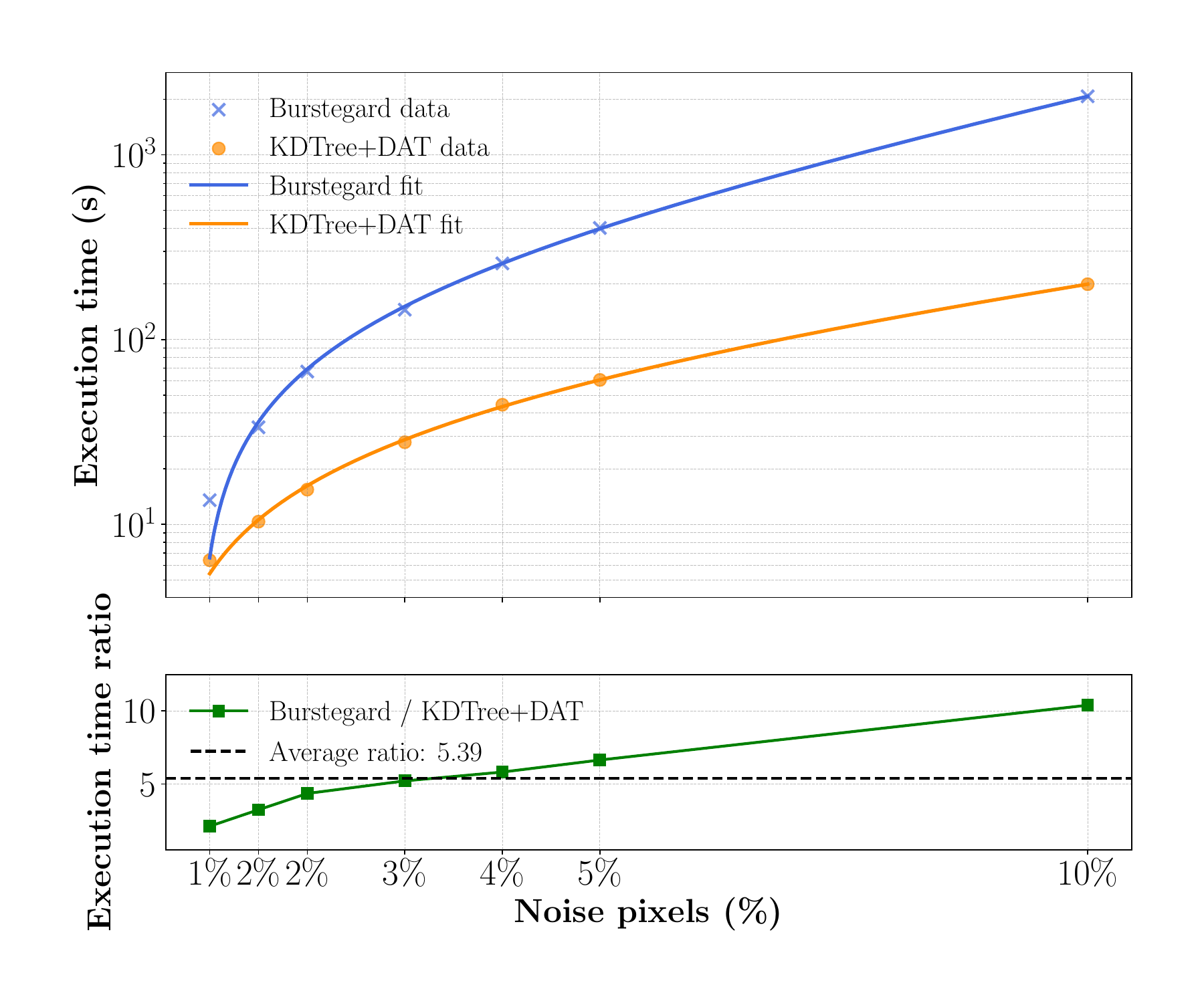}
    \caption{Elapse time of \texttt{burstegard} (blue) and \texttt{KDTree} (orange) algorithms as a function of the fraction of pixels in a $ft$-map. The curves are well approximated by $a ~exp(b\times p)+c$ with $ a = 55.5, b =15.4, c=-59.3$ for \texttt{DAT+KDtree} and $ a = 156, b =26.7, c=-198$ for \texttt{Burstegard}.}
    \label{fig:execution_time_comparison}
\end{figure}

We also compare the elapse time of the clustering algorithms using a $ft$-map spanning \unit[512]{s} and covering frequencies from \unit[22]{Hz} to \unit[2]{kHz}, built with simulated Gaussian noise data, such that results do not depend on the presence of noise transients. We vary the number of pixels with non-zero value such that we keep between 1\% and 10\% of the total number of pixels in the $ft$-map. In Figure~\ref{fig:execution_time_comparison} we show the Wall Clock Time required on a single CPU (Intel\textregistered{} Xeon\textregistered{} CPU E5-2698 v4 @ 2.20\,GHz, utilizing one core) for each algorithm as function of the fraction $p$\, of noise pixels. Both algorithms exhibit similar trends, with elapse time $\Delta t$ well-modeled by an exponential function of the form
\begin{equation}
    \Delta t = a ~ exp{(b \times p)} + c,
\end{equation}
where $a$, $b$, and $c$ are free parameters fitted on the data. 

The elapse time of \texttt{burstegard} is always larger whatever the fraction of pixels, but increases even more rapidily with the number of pixels.
The figure shows that while both algorithms exhibit exponential growth in execution time as the noise level increases, \texttt{DAT+KDTree} scales much slower, especially when the pixel fraction is large. For example, at 10\% noise pixel proportion, the elapse time for \texttt{burstegard} reaches over \unit[1000]{s}, while \texttt{DAT+KDTree} remains under \unit[10]{s}. This indicates that \texttt{DAT+KDTree} is not only faster on average but also significantly more resilient to increased noise level.

On average, \texttt{DAT+KDTree} is $\sim$ 5 times faster than \texttt{burstegard} for pixels fraction comprised between 1\% to 10\%, which is close to the noisiest scenarios observed on real data. As the percentage of noise pixels increases, this ratio grows linearly as shown on Figure~\ref{fig:execution_time_comparison}.
This further underscores \texttt{DAT+KDTree}'s ability to handle higher noise levels more efficiently.

\section{\label{sec:conclusion} Conclusion}

In this work, we presented significant enhancements to the \texttt{PySTAMPAS} pipeline, designed for detecting long-duration transient gravitational-wave signals. We introduced a novel two-stage process for the single-detector analysis phase: first, a Denoising Adaptive Thresholding (\texttt{DAT}) stage utilizing 2D wavelet transforms and statistically-derived adaptive thresholds to clean time-frequency $ft$-maps and select significant pixels; second, an efficient clustering stage employing a \texttt{KDTree} algorithm, a data structure well-suited for fast multi-dimensional searching.

Our comparative analysis, using both simulated signals injected into real LIGO O3b data and noise-only studies, demonstrated the substantial benefits of the \texttt{DAT+KDTree} approach over the standard \texttt{burstegard} clustering algorithm previously used in \texttt{PySTAMPAS}. Key results include a significant reduction in the false alarm rate, particularly for loud triggers, translating to lower detection thresholds for a given significance level. The reduction of low-frequency ($<$\unit[100]{Hz}) triggers is especially interesting as this frequency band is particularly noisy in LIGO-Virgo-KAGRA data. The improved background rejection mainly due to the \texttt{DAT} algorithm contributes to enhanced sensitivity, with detection efficiency improvements observed across a diverse set of astrophysically motivated and ad-hoc waveforms, corresponding to increases in detection reach of up to a factor $2.3$. Furthermore, the new method demonstrated better fidelity in reconstructing signal morphology, especially at the onset and conclusion of signals, while simultaneously showing improved robustness against contamination from instrumental glitches.

A major practical advantage of the \texttt{KDTree} algorithm is its computational efficiency. Our tests showed that the \texttt{DAT+KDTree} pipeline runs significantly faster than \texttt{burstegard}, with speed-ups exceeding a factor of 5 on average for typical noise levels, and exhibiting much better scaling performance as the density of pixels increases. This speed improvement, due to the \texttt{KDTree} algorithm is crucial for timely analysis of large datasets from current and future observing runs. It would also allow to relax some of the parameters such as the minimal number of pixels, to better recover shorter signals ($<$ \unit[10]{s}). Additionally, the \texttt{DAT} stage, based on wavelet analysis, provides effective denoising without the need for prior training on specific signal or noise morphologies, making it broadly applicable and adaptable to varying data conditions and real-time analysis.

Looking ahead, further enhancements could be explored. While the wavelet-based \texttt{DAT} offers advantages in its adaptability, machine learning techniques like Convolutional Neural Networks (CNNs) could potentially offer more sophisticated denoising capabilities, albeit at the cost of requiring representative training data. Beyond clustering, advanced machine learning classifiers, such as gradient boosting algorithms like XGBoost or deep learning models, could be applied to the features of the identified clusters or even directly to $ft$-map regions to further improve the discrimination between true astrophysical signals and terrestrial noise artifacts. Applying this enhanced \texttt{PySTAMPAS} pipeline to the latest data from the LVK detectors promises deeper and more efficient searches for long-duration gravitational-wave transients.

\begin{acknowledgments}
This research has made use of data~\cite{KAGRA:2023pio} or software obtained from the Gravitational Wave Open Science Center (gwosc.org), a service of the LIGO Scientific Collaboration, the Virgo Collaboration, and KAGRA. This material is based upon work supported by NSF's LIGO Laboratory which is a major facility fully funded by the National Science Foundation, as well as the Science and Technology Facilities Council (STFC) of the United Kingdom, the Max-Planck-Society (MPS), and the State of Niedersachsen/Germany for support of the construction of Advanced LIGO and construction and operation of the GEO600 detector. Additional support for Advanced LIGO was provided by the Australian Research Council. Virgo is funded, through the European Gravitational Observatory (EGO), by the French Centre National de Recherche Scientifique (CNRS), the Italian Istituto Nazionale di Fisica Nucleare (INFN) and the Dutch Nikhef, with contributions by institutions from Belgium, Germany, Greece, Hungary, Ireland, Japan, Monaco, Poland, Portugal, Spain. KAGRA is supported by Ministry of Education, Culture, Sports, Science and Technology (MEXT), Japan Society for the Promotion of Science (JSPS) in Japan; National Research Foundation (NRF) and Ministry of Science and ICT (MSIT) in Korea; Academia Sinica (AS) and National Science and Technology Council (NSTC) in Taiwan.

This material is based upon work supported by NSF’s LIGO Laboratory which is a major facility fully funded by the National Science Foundation.
This manuscript was assigned Virgo-Document number VIR-0803A-25 and LIGO-Document number LIGO-P2500509.

\end{acknowledgments}

\bibliography{biblio}

\end{document}